\documentclass[]{eusset}
\usepackage{listings} 

\title{Behind the Feed: A Taxonomy of User-Facing Cues for Algorithmic Transparency in Social Media}

\author{Haoze Guo}
\email{hguo246@wisc.edu}
\affiliation{
  \institution{University of Wisconsin - Madison}
  \country{United States}
}

\author{Ziqi Wei}
\email{zwei232@wisc.edu}
\affiliation{
  \institution{University of Wisconsin - Madison}
  \country{United States}
}

\copyright{}
\doi{10.48340/ecscw2026_014}

\begin{abstract}
People who use social media are increasingly encountering \textbf{UI indicators} that reveal how platforms decide what content to show. However, these transparency cues vary widely across platforms and are often hard to find, making it difficult to compare sites or evaluate whether transparency produces real accountability---or just better user understanding. To address this, we developed a \textbf{classification system} that standardizes how algorithmic presentation is categorized in social media interfaces, including whether platforms explain why content is featured. The system covers three dimensions: \emph{design form}, \emph{information content}, and \emph{user agency}. Applied across six major platforms, it functions as a reference database for identifying recurring UI archetypes and patterns. This framework helps researchers and designers assess whether algorithmic transparency works as intended and supports future interface improvements that increase the \textbf{inspectability}, \textbf{actionability}, and \textbf{contestability} of algorithmic systems.
\end{abstract}


\conferenceName{24th EUSSET Conference on Computer-Supported Cooperative Work}
\conferenceShort{ECSCW}
\conferenceYear{2026}
\submissionType{Posters and Demos}
\journalName{Reports of the European Society for Socially Embedded Technologies}
\issn{2510-2591}

\begin{document}


\section{Introduction}
The development of "algorithmic transparency" on social platforms is being done through the implementation of many small, local interface components that describe how the platform handles recommendations, sponsorships, personalized content, and opportunities to customize content for the user \cite{burrell2016opacity,anannycrawford2018seeing,pasquale2015blackbox}.

Algorithmic Transparency is not one specific area, but it consists of multiple small cues that can be difficult for users to find and offer little assurance to them about the quality of the information. Some of these cues appear when users first interact with a social media platform and explain the provenance of the content that they are engaging with and are located at the front of the user's view; the remaining cues are typically located within menus or linked to other types of documentation. In addition, a small number of the cues are directly connected to the user's ability to take action immediately (e.g., "see less," interest controls) or provide a means to pursue procedural recourse. There are currently gaps within social media interfaces whereby cues can signal “algorithmic transparency” and offer little or no recourse to verify the veracity of the content \cite{anannycrawford2018seeing,bucher2018ifthen}.

We propose an interface-first way to study this gap by introducing a taxonomy of \emph{algorithmic transparency cues}: user-facing UI elements that explicitly reference recommendation/ranking, ad delivery/targeting, or content governance decisions and provide explanation, provenance, or pathways to action. Rather than adjudicating whether explanations are faithful, we treat cues as observable artifacts with design-side properties (placement, interaction cost, specificity/traceability, and agency). This complements evidence that users form ``folk theories'' of curation from limited interface signals and lived experience \cite{radergray2015beliefs,eslami2016folk,devito2018folk}, and supports comparison under interface drift across time, devices, and account contexts \cite{nickerson2013taxonomy,metaxa2021auditing,10.1145/3772363.3798570}.

We structure the paper around three research questions:
\begin{itemize}
  \item{RQ1:} What kinds of algorithmic transparency cues are deployed across major social platforms?
  \item{RQ2:} How do cues vary by design form, information content, and user agency?
  \item{RQ3:} Which transparency functions are systematically underserved?
\end{itemize}

This paper makes three contributions. First, we define \emph{algorithmic transparency cues} as user-facing interface artifacts that disclose, explain, or provide action around platform decisions. Second, we develop a three-part taxonomy for comparing cues across \emph{design form}, \emph{information content}, and \emph{user agency}. Third, we apply this taxonomy to 210 cue instances across six major social platforms, showing that transparency is often present but displaced, weakly verifiable, and unevenly actionable. In particular, recommendation/ranking cues are less often co-present, traceable, or contestable than advertising and governance cues.

\section{Related Work}
This paper is a continuation of research on the ways in which users make inferences about curation from interface encounters, the design of explanations within recommendation systems, and how platforms enact the principle of transparency with respect to governing and advertising. In each of those areas, a single theme remains consistent in being emphasized: While transparency is a characteristic of models, transparency is also an obligation for platforms as they create interfaces that shape transparency through the choice of placement, wording, and potential actions available to users \cite{burrell2016opacity,anannycrawford2018seeing,pasquale2015blackbox}.

Past research has indicated that users rarely gain insights into ranking rationales based solely on formal disclosures. Instead, users form folk theories of how ranking occurs through continuous engagement with the platform and via informal testing with minimal signal from interfaces throughout their day-to-day engagements \cite{radergray2015beliefs,eslami2016folk,devito2018folk}. This has driven researchers to consider and address the significance of looking first at how a platform surfaces explanations and at what expense. Accounts of the opacity of algorithmioic systems assert that the limits to seeing algorithmic systems are physical and sociotechnical cnditions created by institutions and, thus, not only due to missing explanation strings \cite{burrell2016opacity,anannycrawford2018seeing,pasquale2015blackbox}. We approach the research assessment through an empirical lens that considers user-facing cues to be comparable to other artifacts and thus have observable characteristics that include placement, interaction depth, specificity, and temporal traceability.

In recommender-systems research, work on explanatory interfaces has emphasized how explanation form should match user goals and the trade-offs this creates \cite{herlocker2000explaining,tintarev2007explanations,zhangchen2020xairecs}. The HCI literature indicates that transparency interventions may create, rather than destroy, the number of times a user trusts or interacts with the system \cite{kizilcec2016transparency,bucher2018ifthen}. We clarify what the platforms actually provide instead of suggesting new methods of providing explanations, by categorising each of the four forms of explanation (accountability, transparency, and agency) as separate entities \cite{anannycrawford2018seeing,pasquale2015blackbox}.

\section{Definition and Scope}
An algorithmic transparency cue provides a user interface element presented to the end user that outwardly demonstrates the specific outcome of a platform resulting from (a) an algorithmic recommendation or ranking; (b) a delivery of advertisements or location-based targeting; or (c) type of content governance. To qualify as an algorithmic transparency cue, it must provide at least one of the three types of information: Reason/Source Statement, pathway to further information, or actions to take on recourse if necessary \cite{anannycrawford2018seeing,burrell2016opacity}.

Cues will be defined as they appear in an user's perspective and are identified as an observable artifact. For our unit, we will include the (i) in-context feed surfaces including ads/modal menus; (ii) setting/spent dashboards on algorithms that shape an outcome; and, (iii) documentation-like pages found by browsing through an in-context cue. Internal tooling/developer documentation and standard user interface action that have no algorithmic relevance will not be included. As platforms utilise different devices, regions and account states, and apply A/B testing often to change their interfaces, the existence and content of cues will vary both with time and context \cite{metaxa2021auditing,10.1145/3772363.3798570}. We treat integrity- and security-relevant disclosures as part of the governance cue family when they are surfaced to users (e.g., provenance warnings, policy/enforcement notices, or explanations about automated moderation and integrity interventions), especially as social-web content is increasingly reused in downstream AI systems where manipulation of retrieved text is a known threat \cite{guo2026hiddeninplaintextbenchmarksocialwebindirect,guo2026privacyplacebodiagnosingconsent}.

\section{A Taxonomy of Algorithmic Transparency Cues}
Using the above definition to describe each cue we develop a code-able Taxonomy to begin providing a platform-independent comparison of these cues across various platforms \cite{nickerson2013taxonomy,metaxa2021auditing,10.1145/3772363.3798570}. The Taxonomy contains three levels of commitment for the Interface Design Forms for Cues-- \emph{design form} (how the cue is surfaced), \emph{information content} (what the cue asserts), and \emph{user agency} (what users can do in response).

\textbf{Design form} reflects both the discoverability and interaction costs associated with a cue via its Modality and Position (in-feed, overflow menu and settings/help) as well as via its Trigger Mode (or always-on versus user-initiated), Persistence and Interaction Depth \cite{anannycrawford2018seeing,burrell2016opacity}.

\textbf{Information content} reflects both the type of decision a cue supports/recommends (recommendation/ranking, ads, governance), the type of explanation it provides (reason versus provenance/disclosure versus policy framing/mechanism-level) and how specific, traceable (none versus histories/logs versus public repositories), and scoped (item/account/system) \cite{tintarev2007explanations,zhangchen2020xairecs,herlocker2000explaining,anannycrawford2018seeing,pasquale2015blackbox}.

\textbf{User agency} encodes actionability (none; content actions; preference controls; report/appeal), contestability pathway, and feedback-loop visibility (whether consequences are stated) \cite{anannycrawford2018seeing,bucher2018ifthen,burgess2024waist,leerssen2023adarchive,checkfirst2024fulldisclosure}.

Aggregating coded cues yields (i) attribute distributions for cross-platform and decision-type comparison and (ii) synthesis outputs: recurring cue archetypes and a transparency-function gap map spanning legibility, control, verifiability, and contestability \cite{anannycrawford2018seeing,eu2022dsa,dsa_article27}.

\section{Method}
We conduct a qualitative content analysis of algorithmic transparency cues as interface artifacts \cite{krippendorff2018content,neuendorf2017guidebook}. We analyze six major social platforms: Facebook, Instagram, TikTok, YouTube, X, and LinkedIn. We capture cues from the \textbf{mobile applications} by traversing (i) in-feed recommendation surfaces, (ii) post/ad overflow menus and information panels, (iii) ad disclosure and advertiser information flows, and (iv) personalization and governance settings. Each cue instance is captured with screenshots and the full navigation path (interaction steps) required to reach the cue, along with minimal context metadata \cite{metaxa2021auditing}.

Using the taxonomy-derived Codebook, we coded each instance with respect to its design form (Surface Modality/Placement/Trigger/Persistance/Interaction Depth), information content (Decision type/Explanation type/Specificity/Traceability/Scope) and User Agency (Actionability/Pathway to contestability/Feedback loop visibility). We captured ambiguous cases and recorded them in our decision log to promote consistency in the coding process and enhance replicability across and among multiple and ever-changing interfaces \cite{nickerson2013taxonomy,10.1145/3772363.3798570}.

Finally, we organize our findings around the distribution of user agency attributes by decision type and identify recurring combinations of cue attributes. These combinations support a transparency-function gap map spanning legibility, control, verifiability, and contestability \cite{anannycrawford2018seeing,eu2022dsa,dsa_article27}. Because platform interfaces vary by account state, region, device, and A/B testing, we interpret these counts as structured interface snapshots rather than stable prevalence estimates.

\begin{table*}[t]
\centering
\footnotesize
\setlength{\tabcolsep}{4pt}
\renewcommand{\arraystretch}{1.05}
\caption{Directional effects of structural cue choices on transparency functions.}
\label{tab:cue_effects}
\begin{tabular}{p{0.42\textwidth}cccc}
\hline
Structural choice & Legibility & Control & Verifiability & Contestability \\
\hline
Always-on label (provenance only) & $\uparrow$ & $\approx$ & $\downarrow$ & $\downarrow$ \\
Buried in overflow menu & $\downarrow$ & $\downarrow$ & $\approx$ & $\approx$ \\
Routed to documentation portal & $\downarrow$ & $\downarrow$ & $\uparrow/\approx$ & $\approx$ \\
Personalized narrative reason (no trace) & $\uparrow/\approx$ & $\approx$ & $\downarrow$ & $\approx$ \\
Traceability hook (history/repo link) & $\approx$ & $\approx$ & $\uparrow$ & $\approx$ \\
Co-located control (see less/topics) & $\uparrow$ & $\uparrow$ & $\approx$ & $\approx$ \\
Explicit consequence statement & $\uparrow$ & $\uparrow$ & $\approx$ & $\approx$ \\
Explicit report/appeal pathway & $\approx$ & $\approx$ & $\approx$ & $\uparrow$ \\
\hline
\end{tabular}
\end{table*}

\section{Findings}
We report cue ecosystems in measurable interface terms: where cues appeared, what they claimed, and what forms of action or recourse they authorized. The data represents a cross-section of 6 platforms and 210 cue events (identified navigation-captured behavior); the events represent cases in which the cue is used to provide recommendations/rankings, serve advertising and community management/governance/assurance.

\subsection{Dataset Overview and Interaction Cost}
Across 6 platforms, we collected 210 cue instances, corresponding to 74 unique cue types after de-duplication by wording and interaction flow (platform-specific variants retained). Cue instances covered recommendation/ranking (114, 54\%), advertising (67, 32\%), and governance/integrity (29, 14\%) decision types (Table~\ref{tab:dataset}).

We define \emph{interaction depth} as the number of user actions required to reach the first explanatory, disclosure, or action surface associated with a cue. Co-present labels were assigned depth 0 because they were visible at the decision surface. Overflow-menu and information-panel cues had a median depth of 2, settings or dashboard cues had a median depth of 4, and documentation or policy-portal cues had a median depth of 5 (Table~\ref{tab:depth_by_surface}). This access-cost structure matters because it determines whether users encounter transparency at the moment of algorithmic exposure or only through optional navigation away from the decision surface \cite{burrell2016opacity,anannycrawford2018seeing}.

\begin{table}[t]
\centering
\setlength{\tabcolsep}{6pt}
\renewcommand{\arraystretch}{1.12}
\caption{Interaction depth by surface type. Depth is measured as the number of user actions required to reach the first explanation, disclosure, or action surface.}
\label{tab:depth_by_surface}
\begin{tabular}{lr}
\hline\hline
\textbf{Surface type} & \textbf{Median depth} \\
\hline
Co-present label at decision surface & 0 \\
Overflow menu / information panel & 2 \\
Settings / dashboard & 4 \\
Documentation / policy portal & 5 \\
\hline\hline
\end{tabular}
\end{table}

\begin{table}[t]
\centering
\small
\setlength{\tabcolsep}{7pt}
\renewcommand{\arraystretch}{1.18}
\caption{Cue dataset by decision type.}
\label{tab:dataset}
\begin{tabular}{lrrr}
\hline\hline
\textbf{Decision type} & \textbf{Inst.} & \textbf{Types} & \textbf{Platforms} \\
\hline
Recommendation / ranking & 114 & 41 & 6 \\
Advertising & 67 & 23 & 6 \\
Governance / integrity & 29 & 10 & 5 \\
\hline
\textbf{Total} & \textbf{210} & \textbf{74} & \textbf{6} \\
\hline\hline
\end{tabular}
\end{table}

\begin{table}[t]
\centering
\small
\setlength{\tabcolsep}{5pt}
\renewcommand{\arraystretch}{1.12}
\caption{Key accountability attributes by decision type. Recommendation/ranking cues are less often co-present, traceable, or contestable than advertising and governance cues.}
\label{tab:decision_breakdown}
\begin{tabular}{lrrrr}
\hline\hline
\textbf{Decision type} & \textbf{N} & \textbf{Co-present} & \textbf{Trace} & \textbf{Contest} \\
\hline
Recommendation/ranking & 114 & 10 (9\%)  & 8 (7\%)  & 2 (2\%) \\
Advertising            & 67  & 24 (36\%) & 17 (25\%) & 0 (0\%) \\
Governance/integrity   & 29  & 4 (14\%)  & 4 (14\%)  & 9 (31\%) \\
\hline
\textbf{All} & 210 & 38 (18\%) & 29 (14\%) & 11 (5\%) \\
\hline\hline
\end{tabular}
\end{table}

The decision-type breakdown clarifies that the accountability gap is not evenly distributed across platform functions. Recommendation/ranking cues, which shape everyday feed experience, were the least likely to be co-present at the decision surface and rarely included traceability or contestability. Advertising cues were more often co-present and traceable, partly because ad disclosure infrastructures and repositories are more established, while governance/integrity cues were less common but more likely to provide contestability pathways. This motivates the three cross-platform patterns described below: displacement, evidence scarcity, and uneven agency.

\subsection{Three Cross-Platform Patterns: Displacement, Evidence Scarcity, Uneven Agency}
Three patterns recur across platforms and decision types.

\begin{figure}[t]
  \centering
  \includegraphics[width=\columnwidth]{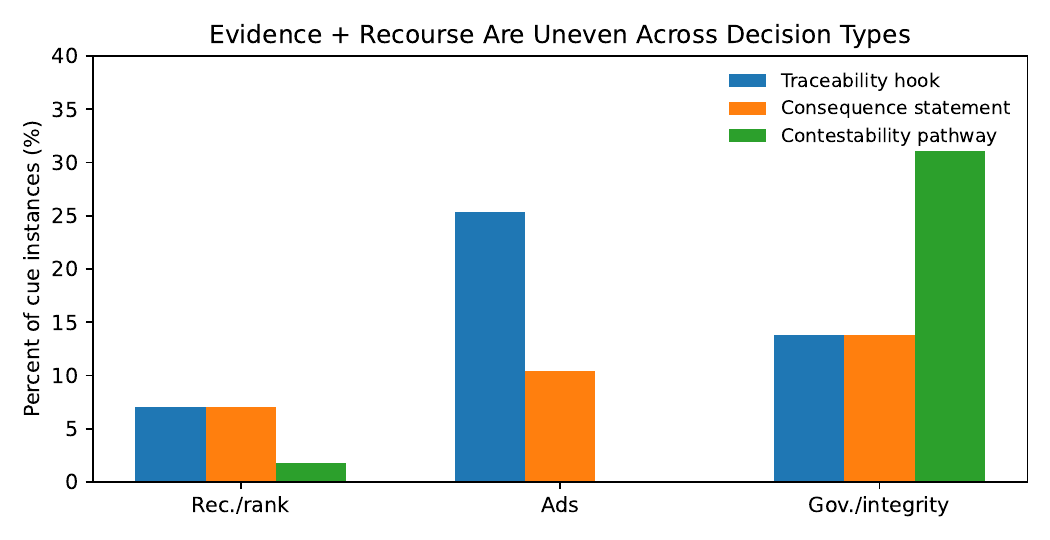}
  \caption{Accountability attributes by decision type. Traceability and contestability are unevenly distributed, concentrating in ads/governance rather than recommendation transparency.}
  \label{fig:attrs}
\end{figure}

\textbf{Displacement is the default, especially for recommendations.}
Only 18\% of cues were co-present at the decision surface, while 49\% required an overflow menu or information panel, 24\% routed to settings dashboards, and 9\% routed to documentation portals. This skew is strongest for recommendation/ranking cues: only 9\% were co-present, compared to 36\% of advertising cues and 14\% of governance cues. In other words, the most common everyday curation context is also where transparency is most often displaced from the moment of algorithmic encounter \cite{burrell2016opacity,anannycrawford2018seeing}.

\textbf{Reasons are common; traceable evidence is rare.}
Across all cues, 46\% were provenance/disclosure-forward and 41\% were reason-giving, but only 14\% included a traceability hook (history/log/repository link) and only 6\% linked to a public repository (e.g., ad libraries) \cite{pasquale2015blackbox,leerssen2023adarchive,checkfirst2024fulldisclosure}. This gap is decision-type dependent: traceability appears more often in advertising (25\%) than recommendation/ranking (7\%), yielding an 18 percentage-point difference. Mechanism-level descriptions were also uncommon (13\%), suggesting that most cues prioritize narrative legibility over checkable evidence.

\textbf{Agency concentrates in ads and governance, not recommendations.}
Overall, 55\% of cues provided no direct action, 29\% enabled content-level actions (hide/see less), and 11\% enabled preference-level actions. Explicit contestability pathways were rare overall (5\%) but concentrated in governance cues (31\% of governance cues), with minimal presence in recommendation/ranking (2\%) \cite{gillespie2018custodians,anannycrawford2018seeing}. Even when controls exist, consequence statements were uncommon: only 9\% of cues clarified scope or duration, meaning many control cues remain “actionable” without making downstream effects inspectable.

\begin{table}[t]
\centering
\small
\setlength{\tabcolsep}{5pt}
\renewcommand{\arraystretch}{1.12}
\caption{Core accountability attributes across all cues.}
\label{tab:attributes}
\begin{tabular}{p{0.62\columnwidth}rr}
\hline\hline
\textbf{Attribute} & \textbf{Count} & \textbf{Prop.} \\
\hline
Co-present at decision surface & 38 & 18\% \\
Requires overflow menu / info panel & 103 & 49\% \\
Routes to settings / dashboard & 50 & 24\% \\
Routes to documentation / policy portal & 19 & 9\% \\
Includes traceability hook (history/log/repo) & 29 & 14\% \\
Includes explicit consequence statement (scope/duration) & 19 & 9\% \\
Provides explicit contestability pathway & 11 & 5\% \\
\hline\hline
\end{tabular}
\end{table}

\subsection{Function Gap: Legibility Outruns Accountability}
Mapping cues onto transparency functions reveals a consistent gap: cues frequently support legibility while underserving verifiability and contestability. In aggregate, 82\% support legibility, while only 14\% support verifiability and 5\% support contestability \cite{anannycrawford2018seeing,pasquale2015blackbox}. The gap is especially stark for recommendation/ranking cues, where 86\% are legibility-forward but only 7\% include traceability hooks and 2\% provide contestability.

A particularly portable “gap” statistic comes from co-occurrence. Among legibility-supporting cues, 83\% provide no verifiability hook; among actionable cues, 86\% provide no consequence statement; and only 9\% of all cues are both co-present and actionable. These combinations clarify why transparency can feel present while remaining difficult to check or contest:
\begin{itemize}
  \item \textbf{Displacement:} 82\% of cues require leaving the decision surface; median interaction depth is 2 actions.
  \item \textbf{Evidence scarcity:} only 14\% include any traceability hook; 6\% link to repositories.
  \item \textbf{Agency imbalance:} contestability is concentrated in governance (31\%) and rare in recommendation transparency (2\%).
\end{itemize}

\section{Discussion}
Algorithmic transparency needs to be reconsidered on the interface level as a commitment—the location of explanations, the types of claims made, and the agency provided. This approach builds upon an ongoing concern within the academic literature regarding transparency \cite{anannycrawford2018seeing,burrell2016opacity,pasquale2015blackbox} —that even if algorithms are considered open, these signals only represent limited opportunities to hold them accountable on the part of the user, and there may also be significant barriers to the establishment, verification, and contestation of such evidence by users. In the same way, we offer our proposed taxonomy of algorithmic accountability as also serving as a threshold for public accountability of algorithms and their everyday application.

\subsection{Design Implications}
The manner in which platforms present transparency and the depth of interaction with it play a critical role in determining whether users encounter transparency as a result of decisions made by the platforms or whether it is reduced to an optional future option. Displacing cues, for example, through the use of overflow menus or documentation portals, adds friction to the user when they need to understand the context of an interaction, while the presentation of co-located cues reduces the user's interaction with the platform and provides the opportunity for greater understanding of the context of their actions \cite{burrell2016opacity,anannycrawford2018seeing}.

At the same time, visibility is not accountability. Platforms often collapse provenance labels, short ``why'' narratives, and preference dashboards into a single transparency story, even though these elements serve different functions. Disaggregating disclosure, explanation, and recourse helps avoid treating minimal provenance signals as meaningful accountability and clarifies when explanation is uncoupled from control \cite{tintarev2007explanations,zhangchen2020xairecs}.

Cues can be differentiated by whether they are explicitly inspectable through claims. Personalized explanations can create feelings of satisfaction; however, alphanumeric forms of user-reported satisfaction will not provide clear evidence about which checking mechanism was used \cite{pasquale2015blackbox,anannycrawford2018seeing,burgess2024waist,leerssen2023adarchive,checkfirst2024fulldisclosure}.

Finally, agency and contestability require more than buttons. Controls such as ``see less'' matter only when users can anticipate consequences (scope, duration, expected effects); otherwise they risk becoming ritual actions without feedback-loop visibility \cite{kizilcec2016transparency,bucher2018ifthen,10.1145/3786995.3787011}. For governance interventions, disclosure without a clear report/appeal pathway is thin accountability, making contestability a distinct surface to design for \cite{gillespie2018custodians,anannycrawford2018seeing}. 

A related design problem is \emph{feedback-loop opacity}: users may be offered controls without being told what those controls will change, for how long, or at what scope. Although 29\% of cues enabled content-level actions and 11\% enabled preference-level controls, only 9\% of all cues included explicit consequence statements. As a result, controls such as ``see less'' or ``hide'' may create a sense of agency without making the feedback loop inspectable. Future transparency cues should pair actions with scope and duration statements, clarifying whether a choice affects one item, a topic, an advertiser, or future recommendations more broadly.

Together, these patterns suggest what we call \emph{compliance by displacement}: platforms can technically provide transparency while routing explanations, evidence, or controls into high-friction paths that reduce their practical availability. In our dataset, only 18\% of cues were co-present at the decision surface, while the remaining cues required additional navigation through menus, dashboards, or documentation portals. This makes interaction depth an auditable property of transparency: not only whether an explanation exists, but whether it is reachable at the moment of algorithmic encounter without substantial navigation cost.

\section{Conclusion}
This paper introduced a taxonomy of algorithmic transparency cues: user-facing interface artifacts that disclose, explain, or provide action around platform decisions involving recommendation/ranking, advertising, and governance. Applying the taxonomy to 210 cue instances across six major social platforms shows that transparency is often present but unevenly accountable. Cues frequently support legibility, yet verifiability and contestability remain limited, especially in recommendation/ranking contexts.

This study focuses on user-facing signals and cannot determine how accurately platform explanations represent the underlying decision processes \cite{burrell2016opacity,doshiivelez2017rigor}. The findings should therefore be interpreted as structured interface snapshots rather than stable prevalence estimates, since platforms vary across region, device, account state, and time. We also do not measure user understanding or behavioral effects \cite{kizilcec2016transparency}. Future work should extend this taxonomy through longitudinal audits that track whether cue depth, co-present share, traceability, consequence statements, and contestability improve or degrade across platform updates.

\bibliography{refs}

\end{document}